\documentclass[preprint,amsmath,superscriptaddress,showpacs,showkeys,amssymb,aps,prb]{revtex4-1}
\usepackage{graphicx,epsfig,multirow,dcolumn,bm}

\begin{document}
\title{Packing fraction related distortion of Mn$_6$C octahedra and its effect on the first order magnetic transition in Mn based antiperovskites}

\author{Aishwarya Mungale}
\affiliation{Department of Physics, Goa University, Taleigao Plateau, Goa 403206 India}
\author{K. R. Priolkar}
\email{krp@unigoa.ac.in}
\affiliation{Department of Physics, Goa University, Taleigao Plateau, Goa 403206 India}

\begin{abstract}
In this paper, we attempt to understand the cause of magnetostructural transformation in Mn-based antiperovskites by calculating EXAFS at the K edges of constituent metal atoms in three antiperovskite compounds, Mn$_3$GaC, Mn$_3$SnC and Mn$_3$InC. These three compounds have very different magnetic ground states despite the similar cubic structure. Our calculations show that the distortions of Mn$_6$C octahedra, which are responsible for the first-order magnetic transition to antiferromagnetic state, depends on the packing fraction of the lattice.
\end{abstract}

\pacs{75.30.Sg; 61.05.cj; 75.30.Kz}
\keywords{Antiperovskites, EXAFS, FEFF, Packing fraction}
\maketitle

\section{Introduction}

The considerable attention received by antiperovskite materials is due to the wide range of properties such as superconductivity \cite{nature,kri}, giant magnetoresistance \cite{goto,Zhang}, magnetostriction effect \cite{Asano}, large magnetocaloric effect and giant negative thermal expansion (NTE) \cite{Takagi,misawa,Iikubo,Huang,Takenaka} demonstrated by them. The properties shown by these materials are associated with a first-order magnetostructural transition from paramagnetic or ferromagnetic to an antiferromagnetic state \cite{Takagi}. The antiperovskite compounds have a chemical formula $M_3M'X$ where $M$ is a 3d transition metal or a rare earth, $M'$ is a metal or metalloid, and $X=$ B, C or N occupying respectively the face centred, $\left(\frac{1}{2},\frac{1}{2},\frac{1}{2}\right)$, the $\left(0,0,0\right)$ and the body centred $\left(\frac{1}{2},\frac{1}{2},0\right)$ positions in a cubic lattice resulting in an inverse perovskite structure. Transition element like Mn occupying the face-centered position results in a triangular arrangement of spins leading to a geometric frustration. In most carbides such as Mn$_3$GaC, the cubic-cubic volume expanding transition results in an interplay of magnetic interactions between the nearest and the next nearest Mn atoms according to the Goodenough-Kanomori rules. Such an interplay of magnetic interactions was believed to stabilize the antiferromagnetic ground state \cite{Fruchart,fru}. Recently, however, extended x-ray absorption fine structure (EXAFS) spectroscopy at the Mn K edge in Mn$_3$GaC established the presence of local structural distortion in the Mn sub-lattice below 250 K. This local distortion lifts the eight-fold degeneracy of the nearest Mn--Mn bonds and gives rise to long and short Mn--Mn bonds. A sudden decrease in the shorter Mn--Mn bond distance, below 170K, results in the first-order ferromagnetic to antiferromagnetic transition in Mn$_3$GaC \cite{elaine2}. This sudden decrease occurs together with a proportionate increase in Ga--Mn and Ga--Ga bond distances resulting in a unit cell volume discontinuity at the magnetic transition.
A change in the M' atom from Ga to Sn resulting in Mn$_3$SnC completely changes the magnetic ground state of the compound. Though both, Mn$_3$GaC and Mn$_3$SnC have a cubic structure with space group Pm$\bar{\mathrm{3}}$m and undergo a cubic-cubic volume expanding transition, their magnetic ground states are different. While Mn$_3$GaC transforms from a ferromagnetic state to an antiferromagnetic state \cite{Takenaka, Elaine-JMMM}, Mn$_3$SnC transforms from a paramagnetic state to a state with complex magnetic order \cite{elaine, Oznur-PRB}. Here too, the Mn EXAFS elucidates presence of local structural distortions in Mn$_3$SnC. The nature of distortions, however, are different than those in Mn$_3$GaC. The role of M' atom in the changes of magnetic ground state of antiperovskite compounds is not clear. Sn is not only larger than Ga in size but also contributes an additional electron to the density of states (DOS) at Fermi level for Mn$_3$SnC. This duality is not resolved by even changing the M' atom to In to realize in Mn$_3$InC. Though In has same number of valence electrons as Ga and Mn$_3$InC has a similar lattice volume as that of Mn$_3$SnC, the ground state of Mn$_3$InC is not analogous to either of them. Mn$_3$InC does not undergo any magnetostructural transition and has only a para to ferromagnetic transition at 430K \cite{Elaine-JAP19}. However, the local structural distortions reveal themselves from the Mn K edge EXAFS studies in the compound.

The structural distortions in the above antiperovskites are local structural in nature and are confined only to the Mn sub-lattice. This nature of structural distortions is evident from the observation of cubic crystal structure for all these three compounds in their high and low-temperature phases. Further, the EXAFS at Ga, Sn and In K edges fit well with the correlations obtained from the observed cubic crystal structure. The origin of such local structural distortions in the Mn sub-lattice of the three antiperovskites is still an open question. In a cubic antiperovskite structure, the nearest Mn--Mn bond distance should be ideally equal to $a/\sqrt2$, where $a$ is the lattice constant. However, EXAFS analysis has shown that, in such cubic antiperovskites containing Ga, Sn or In, compared to the ideal Mn--Mn bond distance, there are longer and shorter Mn--Mn bonds. A comparison of the difference between the long and short Mn--Mn distances reveals a remarkable trend. The difference in long and short Mn--Mn distance is in a sense measure of distortion, and this decreases as the M' atom is changed from Ga to Sn to In. Such a trend cannot be directly related to the lattice volume as Mn$_3$SnC, and Mn$_3$InC have almost the same unit cell volume. Neither can it be associated with the contribution of valence electrons to the DOS as Ga and In have an equal number of electrons in their outermost shell.

To understand the origin of local structural distortions observed in the Mn sub-lattice in the three antiperovskites within their cubic crystal structure and the relation of such local distortions with the M' atom, we have performed ab-inito calculations of EXAFS of Mn and M' atoms (M' = Ga, Sn and In) for the cubic and different locally distorted lattices and compared them with experimental data. The results divulge a close correlation between the local structural distortions and the packing fraction of the cubic unit cell.

\section{Computational details}

EXAFS spectra at the Mn, Ga, Sn, and In K edges in Mn$_3$GaC, Mn$_3$SnC, and Mn$_3$InC have been calculated by employing FEFF 8.4 software based on the self-consistent real-space multiple-scattering formalism \cite{feff,feff2}. For the FEFF calculations, spherical muffin tin potentials were self consistently calculated over a radius of 5\AA. A default overlapping muffin tin potentials and Hedin-Lundqvist exchange correlations were used to calculate x-ray absorption transitions to a fully relaxed final state in the presence of a core hole. Calculations were carried out for Mn K, Ga K, Sn K, and In K edges for the experimentally observed cubic perovskite structure of the respective compound. Besides, different structural distortions centred around Mn atom were also considered by modifying the FEFF input file. In particular, two possible structural distortions were considered and are explained in detail in next section. XAFS was calculated for absorbing atoms occupying the respective crystallographic sites of the perovskite structure. During calculations the amplitude reduction factor, $S_0^2$ was fixed to 0.7 and the $\sigma^2$ for respective paths were calculated using the reported Debye temperatures for the three compounds and the spectrum temperature of 100K. The above-calculated EXAFS spectra for the three samples, Mn$_3$GaC, Mn$_3$SnC and Mn$_3$InC were compared with the experimentally recorded spectra published earlier \cite{elaine2,elaine,Elaine-JAP19}.

\section{Results and Discussion}

\begin{figure}
\begin{center}
\includegraphics[scale=0.7]{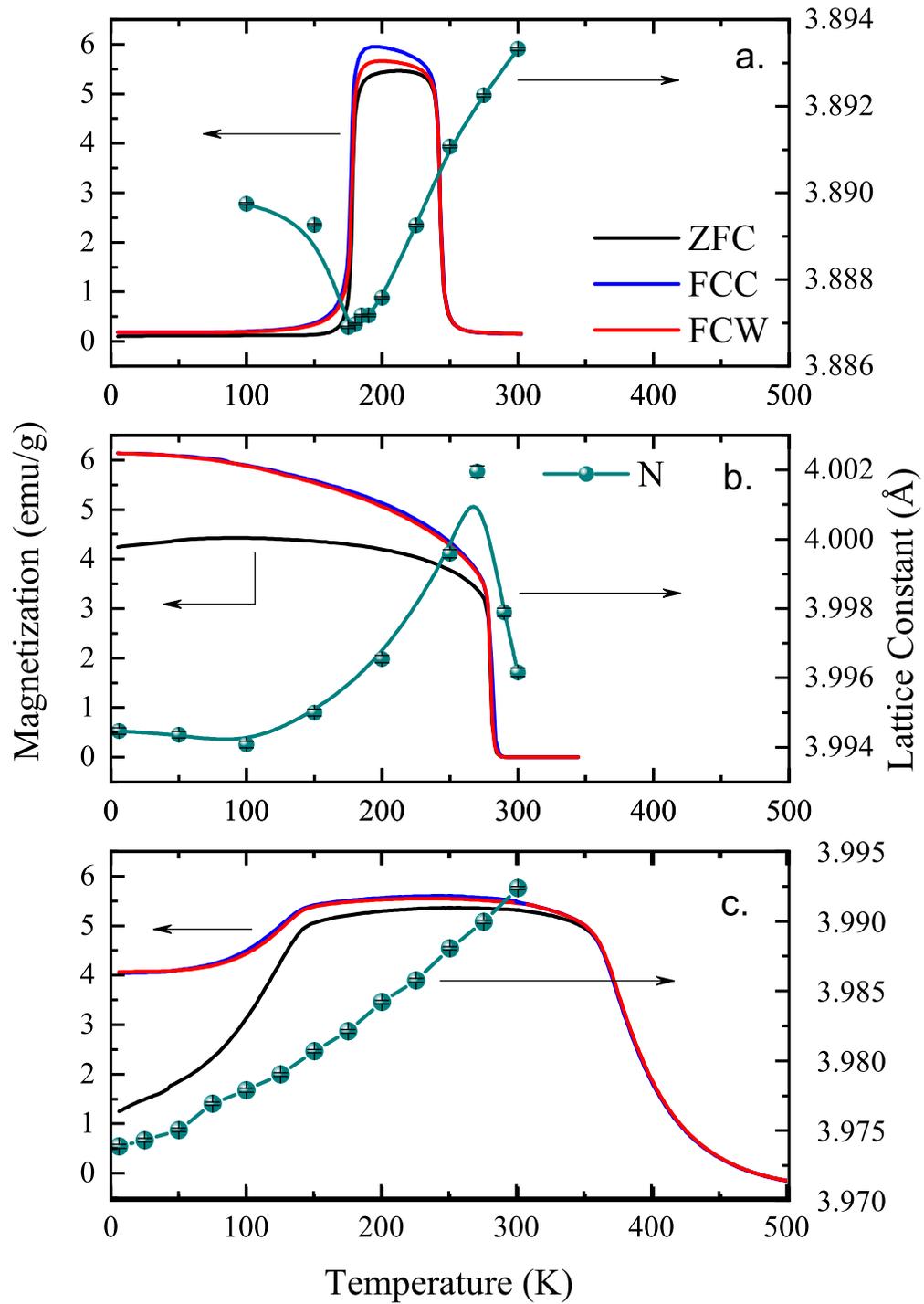}
\caption{Magnetization and variation of lattice parameter as a function of temperature in (a) Mn$_3$GaC, (b) Mn$_3$SnC and (c) Mn$_3$InC}
\label{fig1}
\end{center}
\end{figure}

The three antiperovskite compounds, Mn$_3$GaC, Mn$_3$SnC and Mn$_3$InC show different magnetic ground states despite having similar structures and is exemplified in figure \ref{fig1}. Their lattice constants, magnetic ordering temperatures and the Mn--Mn distances obtained from the analysis of EXAFS data are tabulated in Table \ref{tab1} for clarity. Two of them, Mn$_3$GaC and Mn$_3$SnC exhibit antiferromagnetic ground state even though the magnetic (Mn) atoms are arranged on a triangular lattice. EXAFS studies reported on them ascribe this long-range antiferromagnetic order to local structural distortion of the Mn sub-lattice. It must be pointed out here that though both, Ga and Sn containing antiperovskites are antiferromagnetic the magnetic propagation vector in both the compounds is different. While in Mn$_3$GaC, the spins along (111) direction are antiparallel to each other ($k = \frac{1}{2}, \frac{1}{2}, \frac{1}{2}$); Mn$_3$SnC has a propagation vector $k$ along $\frac{1}{2}, \frac{1}{2}, 0$ direction. On the other hand, Mn$_3$InC does not exhibit antiferromagnetic ground state even though the Mn sub-lattice in Mn$_3$InC is locally distorted.

From Table \ref{tab1} it can also be seen that the separation between the two Mn--Mn bond lengths decreases as the M' atom changes from Ga to Sn  or Ga to In  and exhibits a slight increase for a change in M' atom from Sn to In. Usually such changes are ascribed either to the size of constituent atoms which affects the unit cell volume or to change in contribution of valence electrons to the DOS at the Fermi level. In the present case, however, Mn$_3$SnC and Mn$_3$InC, both have similar unit cell volumes, and Ga and In make an equal contribution to the DOS at Fermi level. So a direct relationship between either the unit cell volume or the valence electron contribution and the magnetic ground state does not exist.

\begin{table*}
\centering
\caption{\label{tab1} Lattice Constants, ferromagnetic transition temperature ($T_C$), magnetostructural transition temperature (T$_{ms}$) and the values of long and short Mn--Mn bond distances and the difference between them ($\Delta$Mn--Mn) in Mn$_3$GaC, Mn$_3$SnC, and Mn$_3$InC}
\begin{tabular}{|c|c|c|c|c|c|c|}
\hline
Compound & Lattice Constant \AA & $T_C$ (K) & $T_{ms}$ (K) & Mn--Mn$_{short}$ \AA & Mn--Mn$_{long}$ \AA & $\Delta$Mn--Mn \AA \\
\hline
Mn$_3$GaC & 3.89936(5) & 242 & 173 & 2.695(8) & 3.105(7) & 0.410(15)\\
Mn$_3$SnC & 3.99672(4) & 274 & 274 & 2.763(4) & 2.867(4) & 0.104(8)\\
Mn$_3$InC & 3.99680(6) & 384 & -- & 2.783(5) & 2.926(14) & 0.143(19)\\
\hline
\end{tabular}
\end{table*}

\begin{figure}
\begin{center}
\includegraphics[width=\columnwidth]{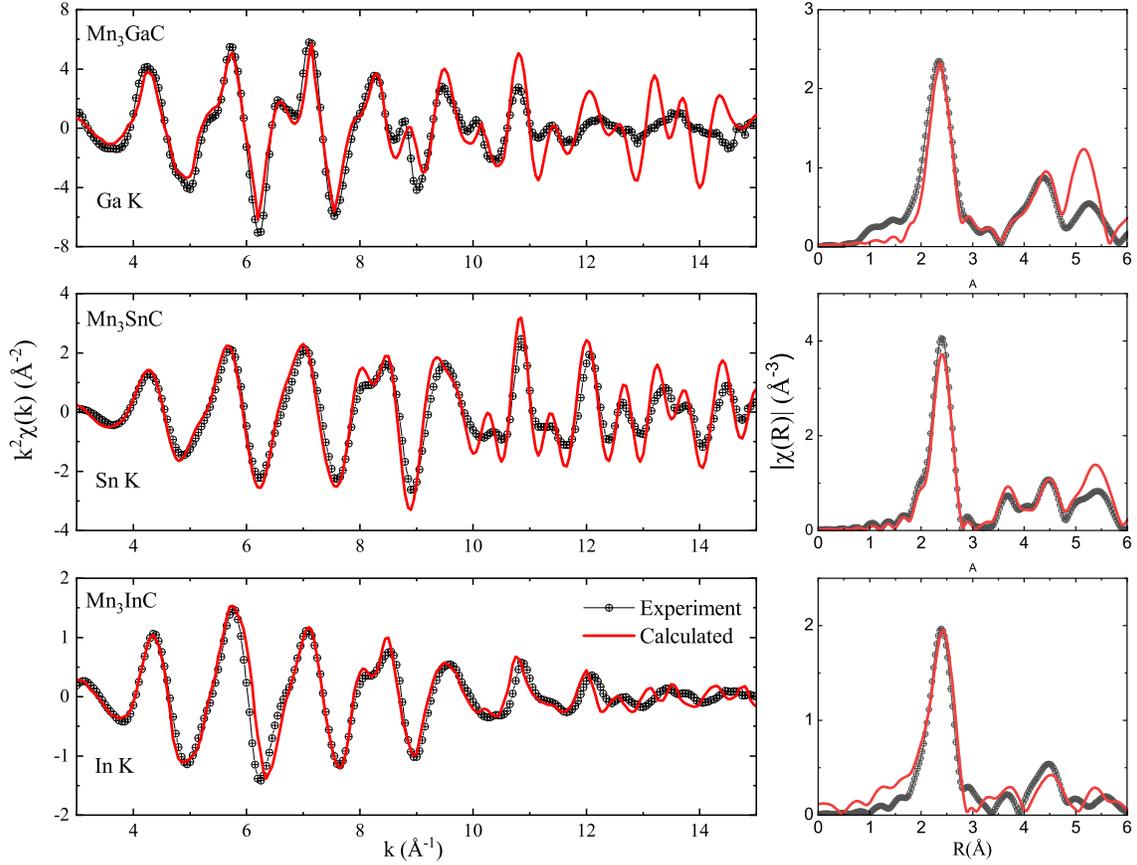}
\caption{Experimentally recorded and FEFF calculated $k^2$ weighted EXAFS and the corresponding Fourier transform at (a) Ga K edge in Mn$_3$GaC, (b) Sn K edge in Mn$_3$SnC and (c) In K edge in Mn$_3$InC}
\label{fig2}
\end{center}
\end{figure}

To understand the possible cause of local structural distortion that is responsible for the observed magnetic properties, EXAFS at the K edges of  Mn and M' (M' = Ga, Sn and In) atoms in the antiperovskite lattice were calculated for different structural models. Two different structural models, (a) based on the cubic crystal structure of the compounds and (b) distorted structural model wherein Mn atoms were displaced from their face centred positions, have been considered. Firstly, EXAFS at the two edges is calculated for the observed cubic crystal structure. The computed results along with the experimentally recorded curves in $k$ space at the M' K edge, where M' = Ga, Sn and In, for the three antiperovskite compounds are presented in figure \ref{fig2}. There seems to be a good agreement between the FEFF calculated and the experimental EXAFS in all three compounds indicating that the local structure around the M' atoms in the antiperovskite compound agrees with the crystal structure of the compound.

\begin{figure}
\begin{center}
\includegraphics[width=\columnwidth]{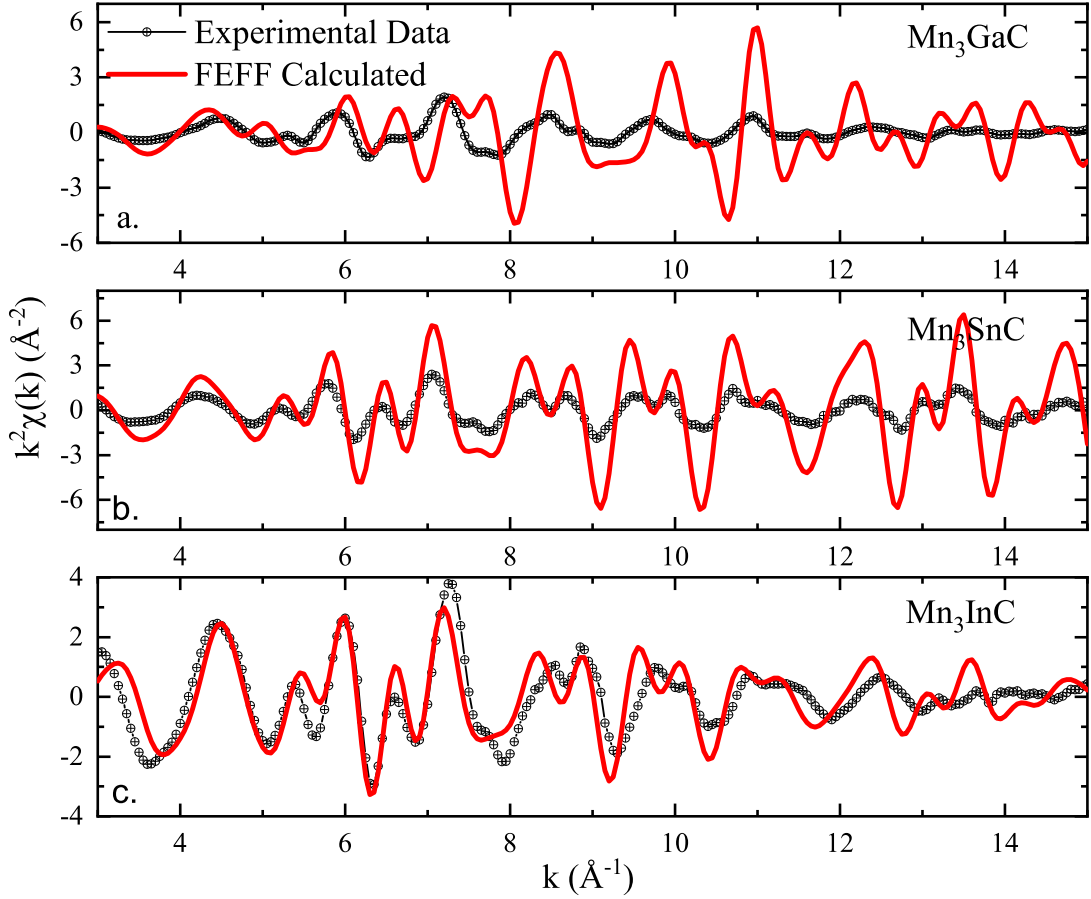}
\caption{Experimentally recorded and FEFF calculated EXAFS based on cubic crystal structure at the Mn K edge in (a) Mn$_3$GaC, (b) Mn$_3$SnC and (c) Mn$_3$InC}
\label{fig3}
\end{center}
\end{figure}

Comparison of experimental and the FEFF calculated $k^2$ weighted Mn K EXAFS as per the crystal structure depicts a completely different story. Figure \ref{fig3}(a) shows the calculated and experimental Mn EXAFS in Mn$_3$GaC. A clear mismatch between the experimental and calculated curves is seen, pointing to the fact that the local structure of Mn is different from that demanded by the cubic crystal structure of the compound. This mismatch is in agreement with the Mn K EXAFS analysis in Mn$_3$GaC which pointed towards a local structural distortion leading to long and short Mn--Mn bond distances. A similar mismatch can be also seen in case of Mn$_3$SnC (figure \ref{fig3}(b)). However, here the mismatch is mostly in the intensity rather than the phase. A phase-matching generally indicates similar local structure. In the case of Mn$_3$SnC, though experimental data analysis has revealed the existence of long and short Mn--Mn bond distances the difference between them is quite small compared to that in Mn$_3$GaC (see Table \ref{tab1}). In the case of Mn$_3$InC the agreement between calculated and experimental EXAFS signal is better than Sn containing antiperovskite compound (figure \ref{fig3}(c)). Mn$_3$InC is the only antiperovskite among the three compounds studied here, that does not display a first-order magnetic transition even though the local structure around Mn is distorted.

As the local structure around M' atoms (Ga/Sn/In) is as per the crystal structure, the distortions around Mn atoms have to be within the cubic unit cell. Further, if one observes the structural correlations around Mn, then the cubic crystal structure demands Mn--Mn and Mn--M' bond distances to be equal. Since the Mn--M' distance can be estimated from the crystal structure, the distortions around Mn atom should be limited to Mn--Mn and Mn--C bonds and hence within a radius of about 3\AA~ from the central Mn atom. Such local distortions could be ascribed to the available free volume within the unit cell. These distortions will not only be limited to Mn--Mn and Mn--C distances but will also vary from compound to compound. The available free volume within a unit cell is related to the packing fraction of the unit cell. The packing fraction can be estimated by dividing the sum of volume of atoms in the unit cell by the unit cell volume. Though such calculation assumes atoms to be hard spheres but allows to estimate of available free volume in each of the three compounds. It is also to be noted that a packing fraction of $\sim$ 74\% implies close packing and in a closely packed lattice, local structural distortions may not be possible without affecting the crystal structurer. Using the reported values of atomic radii \cite{radii} of the constituent atoms for each antiperovskite compound and their unit cell volume reported earlier \cite{Elaine-JMMM,elaine,Elaine-JAP19} the packing fractions were calculated and are reported in Table \ref{tab2}. The values of packing fraction indicate availability of ($\approx$10\%) free volume in Mn$_3$GaC, a little less ($\sim$ 8\%) in Mn$_3$SnC and still further less ($\sim$ 1\%) in Mn$_3$InC.

\begin{table*}
\centering
\caption{\label{tab2} Atomic radius of Mn. Ga, Sn. In and C and lattice volume and packing fraction of Mn$_3$GaC, Mn$_3$SnC and Mn$_3$InC}
\begin{tabular}{|c|c|c|c|c|c|}
\hline
Atom & Atomic Radius & Compound & Lattice Volume & Packing\\
 & \AA &  & \AA$^3$ & Fraction\\
\hline
Mn & 1.27 & Mn$_3$GaC & 59.29 & 63.21\% \\
C & 0.70 & & & \\
Ga & 1.35 & Mn$_3$SnC & 63.84 & 65.16\% \\
Sn & 1.51  & & & \\
In & 1.67 & Mn$_3$InC & 63.85 & 73.12\% \\
\hline
\end{tabular}
\end{table*}

Based on the availability of free volume, distortions were introduced in the local structure of Mn by modifying the FEFF input files. For these modifications we considered two possible scenarios, (a) elongation of Mn$_6$C octahedra along a particular direction while shrinking along the other two directions, and (b) displacement of Mn atoms over a sphere of radius equal to Mn--C bond length. For instance, if the coordinate of Mn atom in the antiperovskite unit cell is $a$/2, $a$/2, 0 (where $a$ is the cubic lattice constant) then the distortion was introduced to move this Mn atom to ($a \pm \Delta a$)/2, ($a \pm \Delta a$)/2, 0 and $\Delta a$ varied from about $0.05a$ to $0.1a$ based on the calculate packing fraction.

Such a choice of inflicting local structural distortions allowed us to consider if Mn--C bonds were also affected due to local structural distortions or were they limited only to Mn--Mn bonds. The choice also has implications on magnetic interactions in the compound. The nearest Mn--Mn bonds are believed to be responsible for ferromagnetic interactions while the next nearest Mn--Mn bonds that are mediated via the C atoms tend to affect the antiferromagnetic interactions. Another scenario proposed to account for the magnetic interactions depends solely on Mn--Mn bonds. Local distortions give rise to long and short Mn--Mn bonds. Here the ferromagnetic interactions are said to arise from the long Mn--Mn bonds while the short Mn--Mn bond aid antiferromagnetic interactions. Thus the choice of structural modifications used here is expected to shed light on the origin of magnetic interactions in these antiperovskite compounds.

\begin{figure}
\begin{center}
\includegraphics[width=\columnwidth]{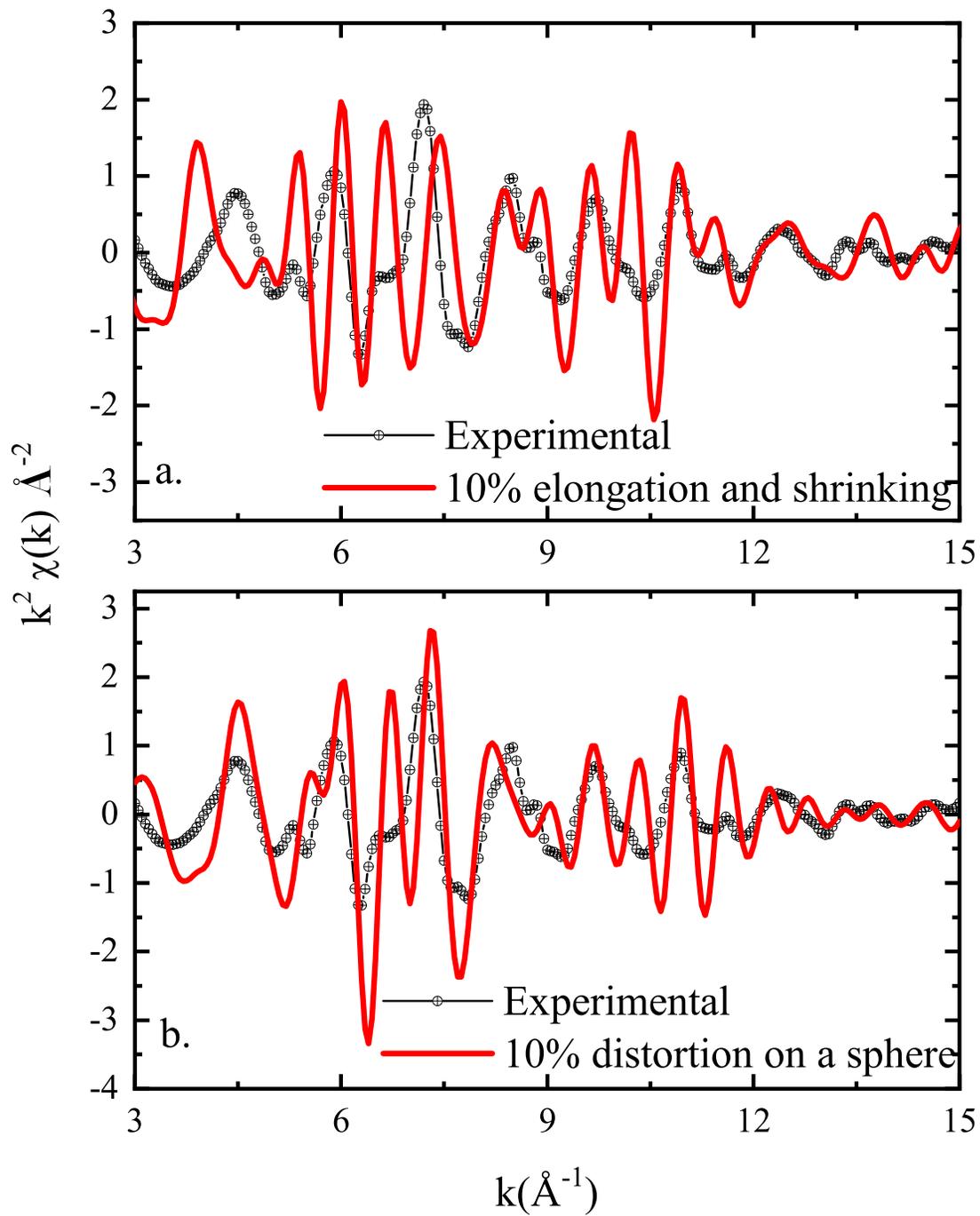}
\caption{FEFF calculated Mn K EXAFS based on locally distorted structural models compared with experimental data in Mn$_3$GaC. }
\label{fig4}
\end{center}
\end{figure}

\begin{figure}
\begin{center}
\includegraphics[width=\columnwidth]{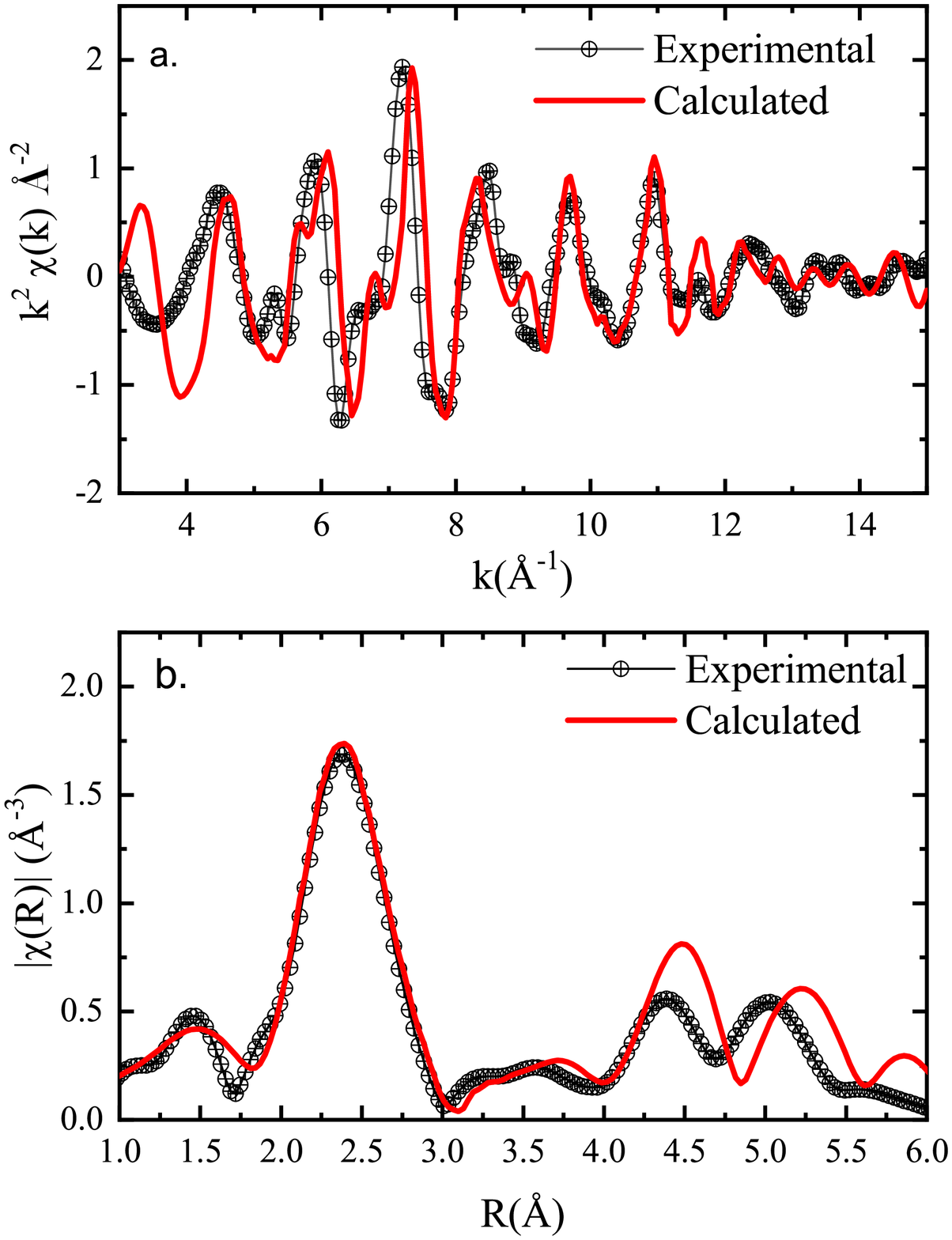}
\caption{Experimental and FEFF calculated Mn K EXAFS based on optimized locally distorted structural model in (a) $k$ space and (b) $R$ space in Mn$_3$GaC.}
\label{fig5}
\end{center}
\end{figure}

\begin{figure}
\begin{center}
\includegraphics[width=\columnwidth]{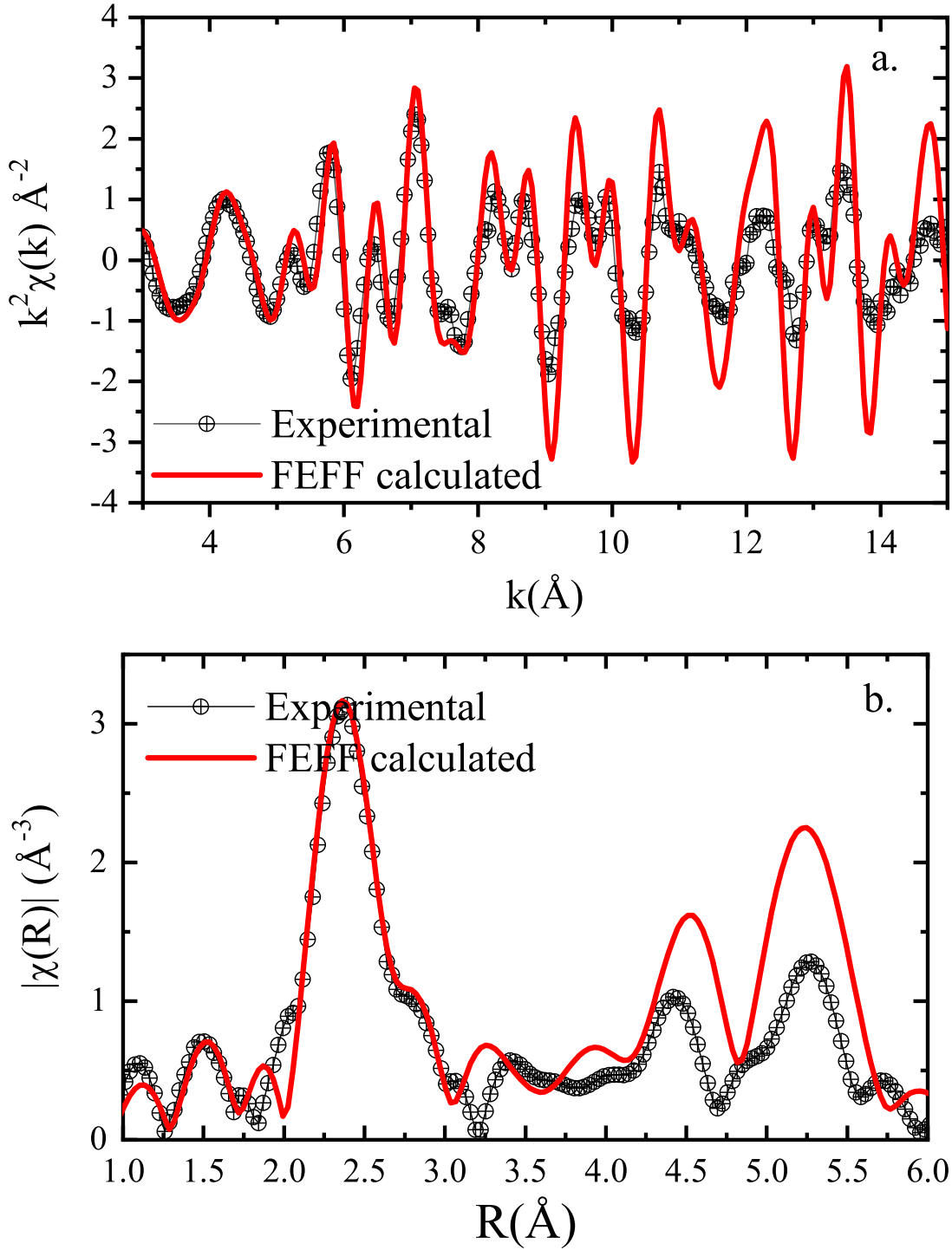}
\caption{Experimental and FEFF calculated Mn K EXAFS based on optimized locally distorted structural model in (a) $k$ space and (b) $R$ space in Mn$_3$SnC.}
\label{fig6}
\end{center}
\end{figure}

In figure \ref{fig4}, the results of calculation of Mn EXAFS in Mn$_3$GaC using the above two scenarios of distortions in the local structure of Mn are presented. While figure \ref{fig4}(a) displays the calculations based on elongation and shrinking of Mn$_6$C octahedra, figure \ref{fig4}(b) enunciates the calculations based on displacement of Mn atoms over a sphere. The calculated curves are compared with the experimental data. It can be readily seen that the calculations of Mn EXAFS based on displacement of Mn atoms over a sphere better replicate the experimental data than the other model. Tweaking of the magnitude of distortion to about 7\% results in a good replication of the experimental data. These results are displayed as Figure \ref{fig5} in $k$ space (Figure \ref{fig5}(a)) and in $R$ space (Figure \ref{fig5}(b)). A similar exercise was carried out for Mn$_3$SnC. Here the distortions were limited to 8\% due to the calculated packing fraction presented in Table \ref{tab2}. It was found that about 5\% distortions reproduce the experimental Mn K EXAFS in Mn$_3$SnC very well and the results are presented in Figure \ref{fig6}. In the case of Mn$_3$InC, the distortions do not seem to improve the agreement between the calculated and the experimental data compared to the Mn EXAFS calculated as per cubic crystal structure indicating the distortions in Mn$_3$InC to be quite small. Such a scenario also augers well with the calculated packing fraction which is just about 1 to 2\% less than close packing value.

Thus the calculations suggest that antiferromagnetic order in cubic antiperovskites is a manifestation of local structural distortions of the Mn$_6$C octahedra. Though the EXAFS recorded at the K edge of M' atom (Ga, Sn or In) can be fitted well with the correlations obtained from the cubic crystal structure, the local symmetry around Mn in the three antiperovskite compounds is not as per the crystal structure. If these two pictures have to coexist, then the local structural distortions of the Mn$_6$C octahedra have to occur within the cubic cage formed by the M' atoms like Ga, Sn or In and are intimately related to packing fraction of the structure. In the case of Mn$_3$GaC, the calculated packing fraction indicates about 10\% of available free space. The experimental EXAFS signal at the Mn K edge can be well replicated by displacing the Mn atoms from their face-centred positions by about 7\% over a sphere of radius equal to Mn--C bond distance. Such a displacement gives rise to long and short Mn--Mn bond distances, keeping Mn--C and the average Mn--Ga bond distance equal to that obtained from the crystal structure. In the case of Mn$_3$SnC, a similar approach with $\sim$ 5\% distortion of Mn$_6$C octahedra replicates the experimental data (see figure \ref{fig6}). The lower distortion in Mn$_3$SnC compared to Mn$_3$GaC augers well with the lower calculated packing fraction (see Table \ref{tab2}). It also matches well with the difference in Mn--Mn$_{long}$ and Mn--Mn$_{short}$ bond distances obtained from experimental EXAFS analysis and reported in Table \ref{tab1}.

The intriguing case, however, is of Mn$_3$InC. Experimental analysis of Mn EXAFS in this compound indicates presence of distortions in Mn$_6$C octahedra. The difference between the long and short Mn--Mn bond distances is similar to that in Mn$_3$SnC. The absence of antiferromagnetism was explained to be due to higher ($>$ 2.75\AA) value of Mn--Mn$_{short}$ bond length. However, the Mn EXAFS in Mn$_3$InC calculated using cubic structure matches well with the experimental data. The calculated packing fraction is also just about 1\% less the close packing value of 74\%. Such lattice packing fraction leaves very little or no scope for distortions in Mn$_6$C octahedra in Mn$_3$InC. The observed distortions in the analysis of experimental EXAFS data is a puzzle. However, careful observation reveals very little difference between the "cubic" fit and the "Delr" fit presented in Ref. \cite{Elaine-JAP19}. Therefore, \emph{ab-initio} calculations of Mn EXAFS based on cubic structure satisfactorily describing the experimental EXAFS in Mn$_3$InC is understandable and could be considered as limitations of the two approaches.

\section{Conclusion}
In conclusion, the antiferromagnetic order in Mn$_3$AC type antiperovskites is mainly due to distortions in the Mn$_6$C octahedra. The extent of these distortions depends on the available free volume within the cubic cage formed by the A site atom. The Mn$_6$C octahedra distort the most ($\sim$ 7\%) in the antiferromagnetic Mn$_3$GaC while minimal distortion exists in ferromagnetic Mn$_3$InC.

\section*{Acknowledgment}
Science and Engineering Research Board (SERB), New Delhi is gratefully acknowledged for financial assistance under the project EMR/2017/001437.


\begin{thebibliography}{0}%
\makeatletter
\providecommand \@ifxundefined [1]{%
 \@ifx{#1\undefined}
}%
\providecommand \@ifnum [1]{%
 \ifnum #1\expandafter \@firstoftwo
 \else \expandafter \@secondoftwo
 \fi
}%
\providecommand \@ifx [1]{%
 \ifx #1\expandafter \@firstoftwo
 \else \expandafter \@secondoftwo
 \fi
}%
\providecommand \natexlab [1]{#1}%
\providecommand \enquote  [1]{``#1''}%
\providecommand \bibnamefont  [1]{#1}%
\providecommand \bibfnamefont [1]{#1}%
\providecommand \citenamefont [1]{#1}%
\providecommand \href@noop [0]{\@secondoftwo}%
\providecommand \href [0]{\begingroup \@sanitize@url \@href}%
\providecommand \@href[1]{\@@startlink{#1}\@@href}%
\providecommand \@@href[1]{\endgroup#1\@@endlink}%
\providecommand \@sanitize@url [0]{\catcode `\\12\catcode `\$12\catcode
  `\&12\catcode `\#12\catcode `\^12\catcode `\_12\catcode `\%12\relax}%
\providecommand \@@startlink[1]{}%
\providecommand \@@endlink[0]{}%
\providecommand \url  [0]{\begingroup\@sanitize@url \@url }%
\providecommand \@url [1]{\endgroup\@href {#1}{\urlprefix }}%
\providecommand \urlprefix  [0]{URL }%
\providecommand \Eprint [0]{\href }%
\providecommand \doibase [0]{http://dx.doi.org/}%
\providecommand \selectlanguage [0]{\@gobble}%
\providecommand \bibinfo  [0]{\@secondoftwo}%
\providecommand \bibfield  [0]{\@secondoftwo}%
\providecommand \translation [1]{[#1]}%
\providecommand \BibitemOpen [0]{}%
\providecommand \bibitemStop [0]{}%
\providecommand \bibitemNoStop [0]{.\EOS\space}%
\providecommand \EOS [0]{\spacefactor3000\relax}%
\providecommand \BibitemShut  [1]{\csname bibitem#1\endcsname}%
\let\auto@bib@innerbib\@empty
\end{thebibliography}%


\begin{thebibliography}{xx}
	\bibitem{nature}T. He et al Nature 411, 54 (2001)
	\bibitem{kri}M. Uehara, T. Yamazaki, T. Kri, T. Kashida, Y. Kimishima and I. Hase, J. Phys. Soc. Japan 76, 034714 (2007)
	\bibitem{goto}K. Kamishima, T. Goto, H. Nakagawa, N. Miura, M. Ohashi, N. Mori, T. Sasaki and T. Kanomata, Phys. Rev. B 63, 024426 (2000)
	\bibitem{Zhang}X. H. Zhang, Y. Yin, Q. Yuan, J. C. Han, Z. H. Zhang, J. K. Jian, J. G. Zhao and B. J. Song, J. Appl. Phys. 115, 123905 (2014)
	\bibitem{Asano}K. Asano, K. Koyama and K. Takenaka, Appl. Phys. Lett. 92, 161909 (2008)
	\bibitem{Takagi}K. Takenaka and H. Takagi, Appl. Phys. Lett. 87, 261902 (2005)
	\bibitem{misawa}K. Takenaka, K. Asano, M. Misawa and H. Takagi, Appl.Phys. Lett. 92, 011927 (2008)
	\bibitem{Iikubo}S. Iikubo, K. Kodama, K. Takenaka, H. Takagi, M. Takigawa and S. Shamoto, Phys. Rev. Lett. 101, 205901 (2008)
	\bibitem{Huang} R. Huang, L. Li, F. Cai, X. Xu and L. Qian, Appl. Phys. Lett. 93, 081902 (2008)
	\bibitem{Takenaka}K. Takenaka and H. Takagi, Appl. Phys. Lett. 94, 131904 (2009)
	\bibitem{Fruchart}J. P. Bouchaud, R. Fruchart, M. Guillot, H. Bartholin and F. Chaise, C. R. Acad. Sci. Paris 261, 655 (1965)
	\bibitem{fru}D. Fruchart, E. F. Bertaut, B. le Clerc, P. Veillet, G. Lorthioir, E.  Fruchart and R. Fruchart, J. Solid State Chem. 8, 182 (1973)
	\bibitem{elaine2}E. T. Dias, K. R. Priolkar, Rajeev Ranjan, A. K. Nigam and S. Emura, J. Appl. Phys. 122, 103906 (2017)
    \bibitem{Elaine-JMMM}E. T. Dias, K. R. Priolkar, and A. K. Nigam, J. Magn. Magn. Mater. 363, 140 (2014)
    \bibitem{elaine}E. T. Dias, K. R. Priolkar, A. Das, G. Aquilanti, Ö. {\c{C}}akir, M. Acet and A. K. Nigam, J. Phys. D: Appl. Phys. 48, 295001 (2015)
	\bibitem{Oznur-PRB}Ö {\c{C}}akir, F. Cugini, M. Solzi, K. R. Priolkar, M. Acet, M. Farle, Phys. Rev. B 96, 014436 (2017)
	\bibitem{Elaine-JAP19}E. T. Dias, A. Das, A. Hoser, S. Emura, A. K. Nigam and K. R. Priolkar, J. Appl. Phys. 125, 063904 (2019)
    \bibitem{feff} A. L. Ankudinov, B. Ravel, J. J. Rehr, and S. D. Conradson, Phys. Rev. B 58, 7565 (1998)
	\bibitem{feff2}J. J. Rehr, J. J. Kas, F. D. Vila, M. Newville "Theory and Analysis of XAFS" in Iwasawa Y., Asakura K., Tada M. (eds) XAFS Techniques for Catalysts, Nanomaterials, and Surfaces. Springer, Cham (2017)
	\bibitem{radii}E. Clementi, D. L. Raimondi, W. P. Reinhardt, J. Chem. Phys. 47, 1300 (1967).
	
	
	
\end{thebibliography}
\end{document}